ARTICLE OPEN

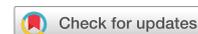

# Analogical discovery of disordered perovskite oxides by crystal structure information hidden in unsupervised material fingerprints

Achintha Ihalage[1] and Yang Hao[1 ✉]

Compositional disorder induces myriad captivating phenomena in perovskites. Target-driven discovery of perovskite solid solutions has been a great challenge due to the analytical complexity introduced by disorder. Here, we demonstrate that an unsupervised deep learning strategy can find fingerprints of disordered materials that embed perovskite formability and underlying crystal structure information by learning only from the chemical composition, manifested in $(A_{1−x}A′_x)BO_3$ and $A(B_{1−x}B′_x)O_3$ formulae. This phenomenon can be capitalized to predict the crystal symmetry of experimental compositions, outperforming several supervised machine learning (ML) algorithms. The educated nature of material fingerprints has led to the conception of analogical materials discovery that facilitates inverse exploration of promising perovskites based on similarity investigation with known materials. The search space of unstudied perovskites is screened from ~600,000 feasible compounds using experimental data powered ML models and automated web mining tools at a 94% success rate. This concept further provides insights on possible phase transitions and computational modelling of complex compositions. The proposed quantitative analysis of materials analogies is expected to bridge the gap between the existing materials literature and the undiscovered terrain.

## INTRODUCTION

Perovskite oxides, identified in $ABO_3$ general form, are utilized in broad spectrum of technological applications. The crystal structure of these materials is conveniently interpreted by an array of corner sharing $BO_6$ octahedra and A-cations sitting at interstitial cages[1] (Fig. 1b). However, in real solids, the perfect periodicity of the lattice is disrupted by several factors. As opposed to the long-standing notion of ordered crystals usually perform better than their disordered counterparts, it is now understood that chemically 'messy' or alloyed perovskite oxides indeed perform exceptionally better in numerous applications. These include ferroelectric photovoltaics[2], electrocatalysts and photocatalysts[3], tunable microwave ferroelectrics with shiftable Curie point[4], solid oxide fuel cells[5,6] and resistive switching memories that are promising to be tailored as memristors and neuromorphic circuit components[7,8]. Thus, perovskite oxides emanate as an imperative class of materials in their own right and carefully engineered disorder can potentially burnish their properties beyond theoretical limits.

Goldschmidt tolerance factor ($t$) and octahedral factor ($\mu$) are simple yet powerful descriptors proposed nearly a century ago to predict the formability of perovskite structure[9]. Recent studies revisited these factors and even developed more accurate descriptors to predict promising ordered perovskites[10–12]. Computational methods of assessing the stability of perovskites include evolutionary algorithms[13], molecular dynamics[12], semi-empirical modelling[14,15] and density functional theory (DFT) calculations of formation enthalpy and energy above convex hull[16]. Nevertheless, adopting high-throughput ab initio methods[17] to model disordered solid solutions with mixed site-occupations is limited due to extreme computational power demanded by large supercells and, several random structure generations and optimizations. Alternatively, machine learning (ML) approaches underpinned by experimental data have been applied to rapidly predict complex functional materials, eventually synthesized to exhibit excellent properties[18–23]. Meanwhile, the complexities brought by compositional disorder such as random atom occupancy of the disordered site, non-stoichiometry and the presence of fractional occupations in the average unit cell have hindered the exploration of the complete composition space for target-driven perovskites discovery. Current supervised ML studies are limited to a minute fraction of disordered perovskites[19,20,24], calling for a general strategy to inversely search the undiscovered territory. Fortunately, unsupervised learning techniques can extract hidden knowledge from raw input features and embed those information in a compressed numerical latent vector[25]. This opens up the avenue of 'analogical materials discovery' where unstudied materials analogous to an existing material are identified by quantitatively comparing their respective latent representations.

In this study, we focus on A-site or B-site alloyed quaternary perovskites, typically formulated as $(A_{1−x}A′_x)BO_3$ and $A(B_{1−x}B′_x)O_3$, respectively. The rules governing A (larger) and B (smaller) cationic radii are well established, yet lenient enough as we identified 49 and 62 periodic table elements that can entirely or partially occupy A-site and B-site, respectively, by incorporating both experimental and theoretical studies[11,12] (Fig. 1a). Such elemental combinations propagate into a staggering and virtually endless diversity of disordered perovskites. Efficient screening of this massive compound pool, first to discover potential perovskites and second, elect promising candidates analogous to a target material/structure (i.e., inverse design) is paramount to supplement the increasing demand for high-performing, inexpensive and eco-friendly materials. By envisaging this objective, we first compiled an all possible candidates pool by capturing the combinations of A and B cations and incrementing $x$ from 0.05 to 0.95 in 0.05 intervals, subject to basic chemistry rules such as charge neutrality and Pauling's valency rule[10,26]. This resulted in a grand total of 591,129 distinct compositions. We then collected an experimental database of $(A_{1-x}A'_x)BO_3$ and $A(B_{1-x}B'_x)O_3$ compositions by exhaustively querying the Inorganic Crystal Structure Database (ICSD)[27]. After discarding the duplicates, we obtained 1758 distinct perovskites and 227 non-perovskites along with their crystal structure information, indicating less than only 0.4% of composition space has been experimentally synthesized before (see Fig. 1c and Supplementary Note 1). This database also comprised double perovskites in the format of $A_2BB'O_6$ and $AA'B_2O_6$, because the composition is the same to the disordered equivalents $A(B_{0.5}B'_{0.5})O_3$ and $(A_{0.5}A'_{0.5})BO_3$, respectively.

[1]School of Electronic Engineering and Computer Science, Queen Mary University of London, London, UK. ✉email: y.hao@qmul.ac.uk

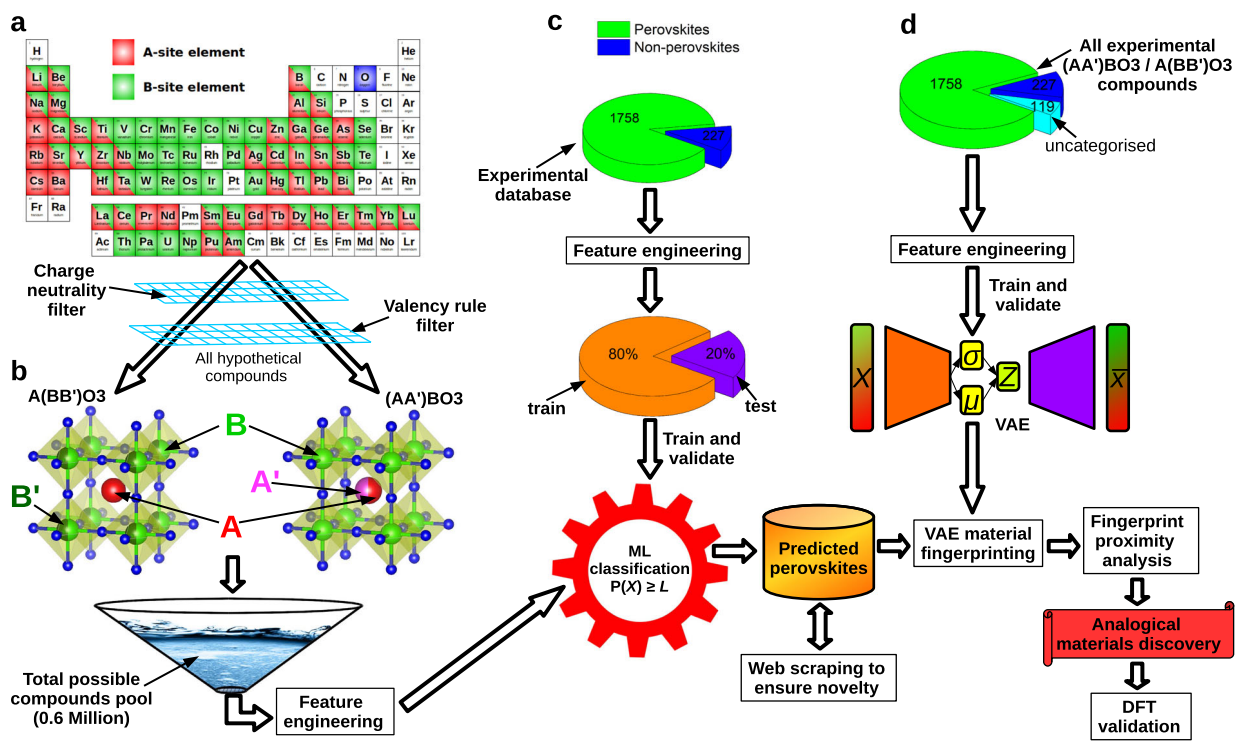

Fig. 1 **Workflow demonstrating the analogical discovery of perovskites. a** Elements that can fully or partially occupy octahedral and interstitial sites. **b** Average unit cells of $(A_{1-x}A'_x)BO_3$ and $A(B_{1-x}B'_x)O_3$ materials, usually realized by X-ray diffraction (XRD) analysis. During total possible chemical formulae generation, Pauling's valency rule is imposed by restricting the mean oxidation state of the A-site ($\bar{n}_A$) to be less than or equal to that of B-site ($\bar{n}_B$); ($\bar{n}_A \leq \bar{n}_B$). We further set the condition, mean ionic radius of A-site ($\bar{R}_A$) should be greater than or equal to that of B-site ($\bar{R}_B$); ($\bar{R}_A \geq \bar{R}_B$), because otherwise the composition relates to interchanged A–B cations. A gradient boosting classifier is iteratively trained on sampled experimental data (**c**), extracted from the ICSD. Total possible compounds pool is classified on a probability threshold (L), in order to improve precision. The originality of positively classified candidates is confirmed by high-throughput web scraping. **d** VAE model is trained on 2104 unlabelled experimental data samples. The euclidean distances between an experimental composition and candidate perovskites are calculated in fingerprint space. Nearest candidates are reported as analogous to the original composition. DFT structure optimization reveals that the analogical candidates preserve the native crystal structure.

An appropriate feature space is developed starting from the chemical formula to represent each material in the experimental database and candidate pool. While characterization equipment such as X-ray photoemission spectroscopy is required to investigate the oxidation state of constituent elements[28], here we implement a heuristic algorithm to unroll fractional oxidation states, closely reflecting the actual cation valences. We further add ten best perovskite descriptors procured from the sure independence screening and sparsifying operator (SISSO) algorithm[11,29] to the feature space, consolidating hand-crafted and data-driven descriptors. A sequential screening strategy that involves ML classification, chemistry informed filtering and automated web scraping is employed to downselect promising unknown perovskites with 93.9% fidelity. We construct a variational autoencoder (VAE) model[30] and train it on a sufficiently large, unlabelled experimental database, sanguinely retrieving a low-dimensional representation of compositions, designated as the material fingerprint (Fig. 1d). Notably, VAEs have been purposed as molecule generative models[31] and feature extractors where the latent vector is sent through another ML model to predict element combinations that will likely form a specific topology, requiring crystallographic data as the input[32]. VAE-obtained latent vector has also been used to predict optical properties of materials[33]. Here, we demonstrate the exclusive potential of material fingerprints learned only from the chemical composition to carry crystal structure and perovskite formability information as numerical vectors, stressing their applicability in analogical materials discovery. Material attributes such as bandgap and phase transition details can be recovered by proximity analysis of fingerprint topology assembled by the VAE. Through a series of statistical testing on crystal symmetry prediction, we illustrate that structurally and physiochemically similar materials likely possess similar fingerprint representations. This phenomenon is employed to discover lead-free analogues of several lead-based ceramics. We further exemplify the versatility of material fingerprints in initializing the DFT simulation of disordered solid solutions by modelling six promising lead-free perovskites (see Fig. 1).

## RESULTS
### Feature engineering
The roadmap to analogical materials discovery begins by establishing a solid and adequate collection of candidate compounds for the parent experimental material or class of materials. We use pymatgen[34], an open-source python library for analyzing the materials in our database. Pymatgen is equipped to guess the most probable oxidation states of elements in a composition based on ICSD statistics. In the case of mixed valence constituent element, this may lead to a fractional oxidation state that must be unfolded to estimate the effective ionic radius of that element and the associated crystallographic site in the average unit cell. Considering the smallest possible supercell that relates to the integer formula, our algorithm first tries to solve for the valence states of individual ions by exploiting the common oxidation states of the mixed-valence element in question. If no solution is found, it looks through other less common oxidations states. Whenever the algorithm finds more than one configuration of valence states can result in the same fractional oxidation state, the one with the smallest oxidation state numbers and the lowest variation in ionic radii is selected (see Supplementary Note 2). We expect the perovskite structure is more likely to be retained under these conditions because, if there is large difference in size between the constituent cations, the octahedral units share faces, distorting the perovskite lattice more[6]. However, it should be noted that mixed valence states are highly dependent on the element, composition, vacancies etc., requiring experimental characterization for precise investigation. Our algorithm provides a systematic estimation of the mean ionic radius of such elements, permitting us to calculate $t$, $\mu$ and develop the feature space. Subsequently, $t$ and $\mu$ of $(A_{1-x}A'_x)BO_3$ type compositions are computed as[1];

$$t = \frac{[(1-x)R_A + xR_{A'} + R_O]}{\sqrt{2}(R_B + R_O)} \quad (1)$$

$$\mu = R_B/R_O \quad (2)$$

Similarly, for $A(B_{1-x}B'_x)O_3$ type compositions;

$$t = \frac{(R_A + R_O)}{\sqrt{2}[(1-x)R_B + xR_{B'} + R_O]} \quad (3)$$

$$\mu = [(1-x)R_B + xR_{B'}]/R_O \quad (4)$$

where $R_A$, $R_{A'}$, $R_B$, $R_{B'}$ and $R_O$ are Shannon's ionic radii[35] of A, A′, B, B′ cations and oxygen anion, respectively. Mean ionic radius is used for mixed valence elements. The feature space is composed of 55 elemental and 11 compositional features, 10 SISSO descriptors and 5 feature combinations including $t$ and $\mu$ (provided in Supplementary Table 1). SISSO is an efficient, data-driven screening approach that finds near-optimal material descriptors relating to a property of interest from a colossal number of potential descriptors obtained by applying a set of mathematical operations to one or many primary features[29]. In our specific case, we have 71 primary features, yielding a total of ~$10^{71}$ SISSO descriptors. By training the SISSO algorithm on around 500 randomly sampled compositions from our database and considering a subspace size of 1000, we extracted 10 best descriptors reflecting perovskite formability (Supplementary Note 3). The features related to oxygen are common to all compositions (i.e., no relative importance) and have thus been omitted when mapping the feature space into perovskite or non-perovskite using ML.

## Quaternary compositions classification

We examined the performance of four ML algorithms in the present classification task to select the best model to proceed. These include decision tree (DT), random forest (RF), support vector machine (SVM) and gradient boosting classifier (GBC). In order to assess the performance metrics, we iteratively trained and evaluated the individual models on sub-databases generated by integrating 227 non-perovskites with 250 randomly sampled perovskites from 1758 total perovskites. Such a strategy is required to maintain the class-balance of individual training sets, minimizing the potential overfitting to a particular class. It also ensures that all available perovskites are exploited without being dissipated. All four algorithms are trained and validated for 100 independent iterations by splitting each sub-database into 80% training and 20% test sets[19]. Table 1 summarizes tenfold cross validation (CV) results. GBC is selected to proceed as it provided the best performance with 93.91% (±0.96%) and 93.33% (±2.26%) tenfold CV and test mean classification accuracies, respectively. More details about these metrics including hyperparameter tuning are available in Methods, and Supplementary Fig. 1 and Supplementary Note 4. Figure 2b,c depict confusion matrix and receiver operating characteristic (ROC) curves, respectively, as obtained by tenfold CV. Moreover, five SISSO descriptors make it to the top ten features ranked by their relative importance (Fig. 2a). However, we note that the GBC model achieves 93.71% and 93.16% tenfold CV and test mean accuracies, respectively, even in the absence of SISSO descriptors in the feature space, endorsing the fact that perovskite formability is heavily dependent on geometrical features such as cationic radii.

We then classify the total possible compounds pool (0.6 Million candidates) by following the same procedure of 100 iterations, except that, this time, each GBC model is trained on the full corresponding sub-database without train/test splitting. Figure 2d highlights the geometric region that most probably results in a stable perovskite structure. Moreover, B-site doping is a prevalent technique used to enhance and tune the properties of many perovskites. Figure 3 shows the classification map of such systems for commonly observed A and B cations. Nonetheless, our objective here is to screen down a huge composition space to a sufficient collection of potential perovskites, such that the materials will have a greater chance of potentially being synthesized in single phase perovskite structure. Therefore, we extract the compositions that have over 95% mean perovskite likelihood and, that were classified 100 out of 100 times into the perovskite category, as the most promising perovskites[19]. This increases the precision (how many selected items are relevant) of the predicted set at the cost of some possible perovskites not being selected. This tactic is also expected to address the ongoing criticism that ML models tend to predict a high rate of materials to be stable, which are actually not[36]. Likewise, 46,228 highly-probable candidate perovskites are screened, narrowing the composition space by 92%. We further downselect the compositions made up of elements in their common oxidation states. This was followed by removing all the compositions whose quaternary system is available in the ICSD relating to our format as already studied materials. For example, when we find $Pb(Zr_{0.5}Ti_{0.5})O_3$ is present in the ICSD, we remove all $Pb(Zr_{1-x}Ti_x)O_3$ compositions that belong to the quaternary system Pb-Zr-Ti-O. Furthermore, an automated web scraping tool is implemented to search for any mentioning of the resulting materials/systems on the internet, which were removed in turn. The competency of this tool is tested by scraping 576 $ABX_3$-type experimental materials reported in ref. [11]. It found the existence of 562 compounds on the web indicating over 97% success rate. This multi-step, novelty-ensuring screening flow further decreased the perovskite candidate headcount to 10,790, which will serve as the search space for analogical discovery of perovskites.

| Table 1. | Performance metrics obtained by tenfold CV for perovskites classification. | | | |
|---|---|---|---|---|
| | GBC | RF | DT | SVM |
| Accuracy | 0.939 (±0.009) | 0.907 (±0.007) | 0.917 (±0.013) | 0.924 (±0.007) |
| Precision | 0.947 (±0.019) | 0.902 (±0.092) | 0.943 (±0.003) | 0.917 (±0.035) |
| Recall | 0.946 (±0.021) | 0.881 (±0.114) | 0.943 (±0.003) | 0.914 (±0.042) |
| F1-score | 0.946 (±0.001) | 0.879 (±0.013) | 0.943 (±0.001) | 0.913 (±0.003) |

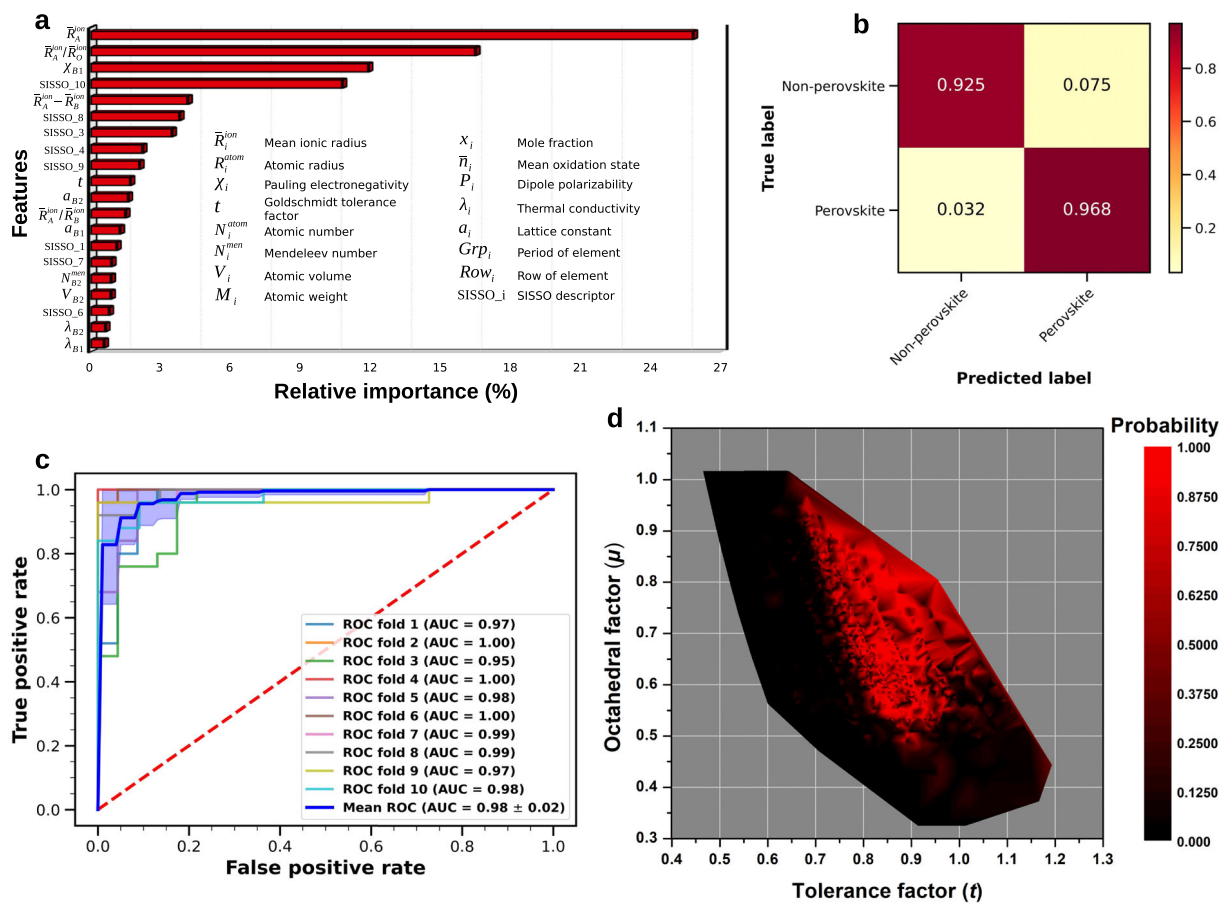

**Fig. 2 GBC classification metrics and prediction results. a** Relative importance of top 20 features. Notably, tolerance factor ($t$) has dropped to the 10th place, which may indicate that $t$ alone is not sufficient to predict the perovskite formability of complex solid solutions at a good accuracy. **b** Normalized confusion matrix obtained by tenfold CV. **c** ROC curves obtained for each fold in tenfold CV and the mean ROC curve. Area under the curve (AUC) summarizes the performance of each GBC into a single measure where high AUC indicates the model has a greater capability of distinguishing between classes. This demonstrates that our model performs consistently even with different random data divisions. Shaded region shows the standard deviation of mean ROC curve. **d** Classification probability heatmap for 0.6 Million compounds pool, plotted as $\mu$ vs $t$.

### Unsupervised material fingerprinting

Learning numerical vector representations of materials, that is fingerprinting them, based only on the features extracted from their chemical formulae facilitates the similarity investigation of two or more materials by geometrical distance arithmetics between those vectors. By design, the compositions that are chemically and structurally alike should fall close to each other and disparate compositions should be further apart in the fingerprint space. An autoencoder neural network, composed of an encoder that efficiently compresses discrete material features into a latent vector and a decoder that learns to reconstruct the same features starting from those encodings, is an unsupervised technique of retrieving the material fingerprint, which refers to the encoding itself. However, a fundamental problem of such a strategy is the possibility of forming blind spots in the latent space, which, when decoded, may have no relevance to a real material. Variational autoencoders (VAEs), in contrast to vanilla autoencoders, resolve this issue by enforcing the encoder to produce two vectors relating to a set of means $\mu$ and standard deviations $\sigma$ that closely resemble a standard Gaussian distribution ($Z \sim N(0, 1)$), which is then sampled to obtain the latent vector, causing the latent space to be continuous (Fig. 4a). Intuitively, this means that the latent encoding of a material is not unique, but stochastically distributed ($Z \sim N(\mu, \sigma^2)$) around the centre determined by the mean vector, enabling the decoder to 'see' the whole latent space during training and driving the encoder to generate interpretable representations. While the robustness of VAEs lies in the stochastic nature of the latent embedding, this vector is not adoptable as the material fingerprint due to its non-uniqueness. Therefore, we designate the mean vector ($\mu$) produced by the encoder as the material fingerprint. In our VAE architecture, $\mu$, $\sigma$ and latent vector all are two-dimensional (2D), yielding 2D numerical fingerprints (see Methods for more details on VAE implementation). We further discuss other dimensionality reduction algorithms in Supplementary Note 5, namely, vanilla autoencoder (Supplementary Fig. 3), principle component analysis (Supplementary Fig. 4) and t-SNE (Supplementary Fig. 5).

Owing to the fact that training VAEs is unsupervised, that is no labelling of compositions is required, we used all available $(A_{1-x}A'_x)BO_3$ and $A(B_{1-x}B'_x)O_3$ compositions totalling 2104 to train and validate the model. Over 90% of experimental data relate to ambient conditions, so do our predictions. The database included 119 materials that were difficult to be reliably deposited into either of perovskite or non-perovskite categories. Clearly, there is no hard boundary between these two classes, separating them perfectly. Ultimately, it comes down to which degree of distortion from the ideal cubic perovskite lattice can be tolerated to still regard the resulting structure as perovskite? We let the VAE model to place the materials in the region that they 'deserve'. The input to the VAE contains the same features as used in the previous classification step. Despite not being explicitly trained as perovskite or non-perovskite, the material fingerprints capture this information and form separate regions in the fingerprint space as can be observed from Fig. 4b. In addition, the alloyed site is clearly recognizable from the 2D fingerprint. Notably, some of the non-perovskite-labelled materials fall within the perovskite region as circled in Fig. 4b. Instead of disregarding them as ambiguous, we scrutinized the root cause for this and found that at least 14 out of 20 materials falling inside the circle were described with the terms 'perovskite' or 'hexagonal perovskite' in other areas of the literature (see Supplementary Table 2). Apparently, these compounds are more 'perovskite' than 'non-perovskite'. This is a prime example to elucidate that the materials are fingerprinted on their own merits. It also draws the attention to what appears to be outlier data for cross checking the correctness of data labelling.

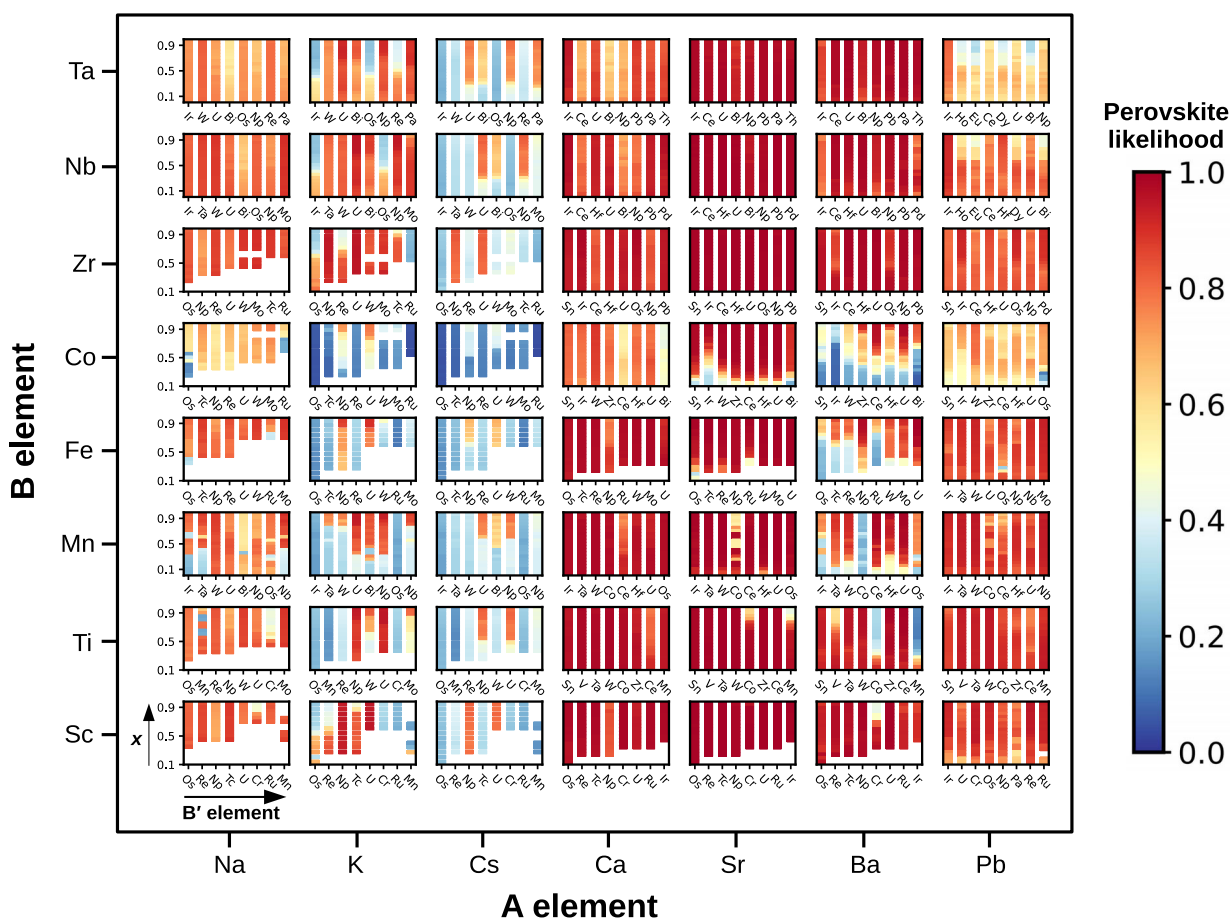

**Fig. 3 Perovskite likelihood of selected $A(B_{1-x}B'_x)O_3$ systems as predicted by the GBC model.** Each panel depicts the classification probability variation as the content $x$ of different $B'$ cations increases, for some commonly observed A and B element combinations. Empty spaces imply no charge balanced probable compositions exist. Caesium cation is too large to occupy A-site and the formation of a stable perovskite is generally improbable. ICSD reports one quaternary perovksite oxide, $K(Ta_{0.77}Nb_{0.23})O_3$, with potassium fully occupying A-site, as opposed to four stable ternary perovskites, $KNbO_3$, $KTaO_3$, $KUO_3$ and $KIO_3$. Despite having no training data with A-site K and B-site U occupied perovskites, the model predicts that U can be alloyed with Nb, Ta or even Ti and Zr to forge a perovskite. This claim is somewhat supported by above ternary oxides. Supplementary Fig. 2 shows the composition map where A-site is partially occupied by K. Strontium, an ideally sized cation sitting at A-site can form plenty of stable perovskites, supporting diverse B-site substitutions for tailoring the properties of materials. Alkaline earth metals in A-site generally show high perovskite likelihood. Likewise, these elements can replace toxic lead in many useful perovskites.

Now, the cardinal question is, are these fingerprint vectors really learned such that they embed the underlying chemical and structural properties of materials? To answer this question, we investigated small local clusters in the fingerprint space. The key idea is that the unsupervised model finds similar fingerprint representations for analogous compositions, which can be corroborated with a k-nearest neighbours approach. Anticipating that similar materials should find themselves close to each other, we compared the crystal system (cubic, orthorhombic, tetragonal, etc.) of each material with that of the five nearest neighbours (NNs) in the 2D fingerprint space. This can be viewed as a leave one out cross validation (LOOCV) technique, because, despite never being trained to predict the crystal system, all the other materials except the one being tested are present on the fingerprint space. In order to quantify materials analogies in native domain, we define euclidean distance derived similarity measure ($S_E$) as; $S_E(p, q) = 1/(1 + d_E(p, q))$ where $p$ and $q$ are two composition points and $d_E(p, q)$ is the euclidean distance between $p$ and $q$. If majority of neighbours vote for (belong to) the same crystal system, that will be assigned as the predicted crystal system of the original composition. In the case of equal number of votes, the one relating to the lowest euclidean distance is selected. By performing this analysis for all 2104 materials, we found that the 5-NNs can determine the correct experimental crystal system of the parent composition with a success rate of 71.8%. This implies that the unsupervised material fingerprints indeed carry crystal structure information concealed in its relative locality. Note that the optimal number of neighbours is found to be 5 by heuristic tuning. Figure 4c illustrates crystal system prediction performance as a normalized confusion matrix. Moreover, this strategy is extensible to predicting the space group of compositions, which resulted in 59.9% prediction accuracy (see Supplementary Fig. 6 for confusion matrix). Given that the total number of distinct space groups present in our database is 66 and the model is not explicitly trained to predict them, nearly 60% accuracy figure reflects that even such diminutive information is inherent in the material fingerprint. Although the data circumstances are different, material fingerprint based prediction of crystal system proposed here offers around 6% accuracy improvement over existing multi-class supervised approaches that rely on multitude of magpie features obtained only from the composition[24,37]. Furthermore, space group detection of experimental compounds, even by training a supervised convolutional neural network (CNN) on about 150,000 powder XRD patterns is proved to be challenging, with a reported accuracy of 81%[38]. With no supplemental knowledge on crystallographic or XRD data of complex solid solutions, our model performs appreciably close to the state-of-the-art CNNs on this aspect, albeit the differences between sample sizes and data contexts should be accounted.

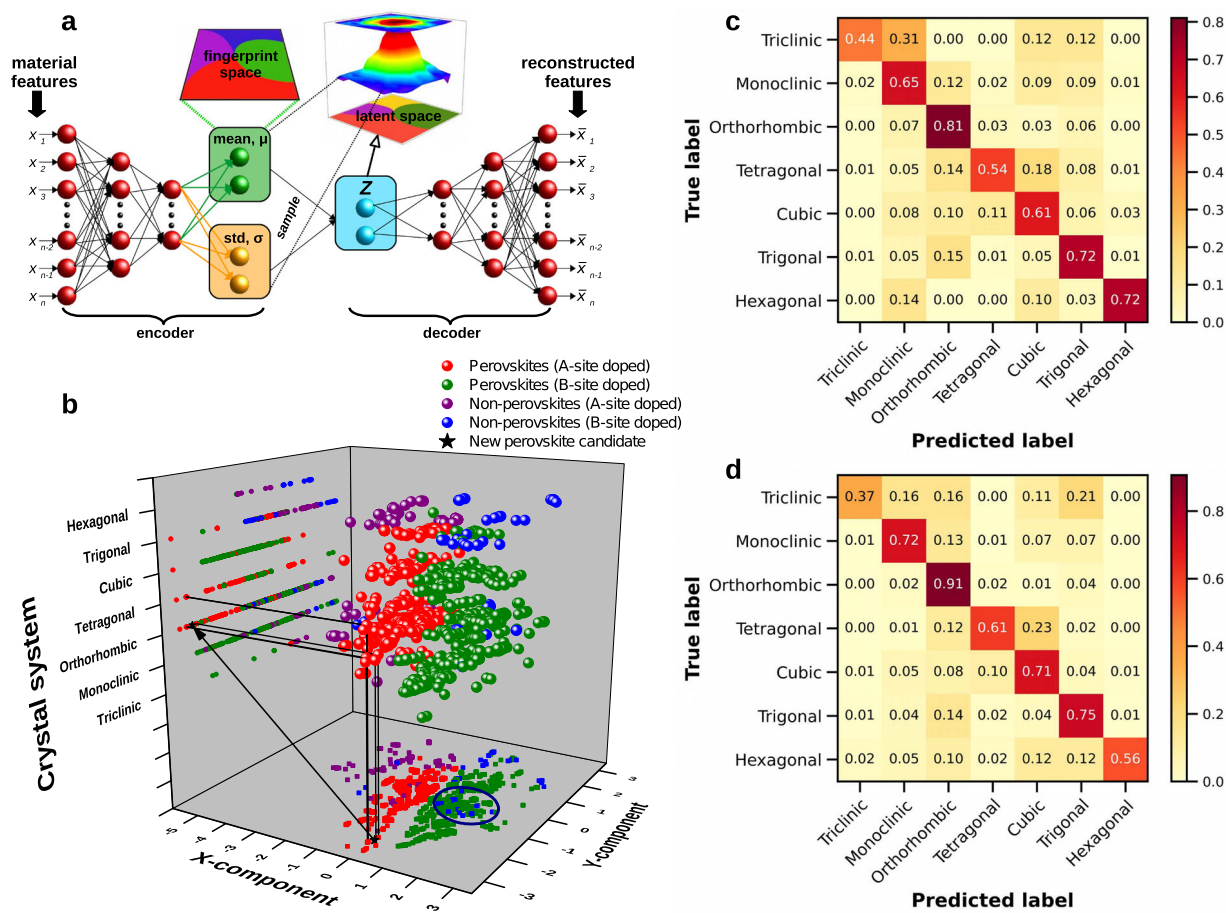

**Fig. 4 Unsupervised material figerprinting using VAE. a** VAE architecture. Encoder must find an efficient compression, discarding bulk of irrelevant information from the input to support the decoder to reconstruct the input itself from a highly compressed dimension. Because latent vector is sampled from a probability distribution, during the training process, the decoder learns not only a single point in latent space is relevant to a particular material, but also the nearby points. This removes blind spots to the decoder and enables placing unseen compositions on the fingerprint map. **b** Crystal system vs 2D fingerprint visualization of the experimental materials. XY plane relates to the fingerprint space. 14 non-perovskite labelled materials out of 20 landing inside the circled area were identified as 'perovskite' or 'hexagonal perovskite' in separate literature, which may not have been examined during data labelling. The crystal system of the indicated candidate is predicted to be orthorhombic based on the plurality vote of 5-NNs. **c** and **d** display the normalized confusion matrices detailing the crystal system prediction of 2104 experimental materials as obtained by proximity analysis of fingerprints and supervised gradient boosting classifier, respectively.

To assess how well this fingerprint proximity oriented approach contrasts with supervised parametric models, we evaluated the crystal symmetry detection performance of GBC, DT, RF and SVM classifiers using LOOCV for unbiased comparison. The feature space is chosen to be the same as used in perovskites classification step. The performance metrics are summarized in Table 2. GBC achieves the highest crystal system classification accuracy of 78.8% (see Fig. 4d) while the proposed scheme, relying only on the 2D fingerprints, outperforms supervised algorithms such as SVM and RF. The trend is similar for space group classification. Most importantly, unsupervised deep learning techniques are quite rigorous when finding their own representations, and advantageous in inversely exploring the compositional space, which is not feasible in supervised ML regime. For instance, when the database is of sufficient size and diversity, the representations learned by VAE form an information-embedded topology that can be used not only to predict the crystal structure of unknown compositions but also to discover compositions analogous to target experimental materials, a point that we come back later in this paper.

| Table 2. | LOOCV performance metrics of crystal system and space group classification. | | | | | | | | | |
|---|---|---|---|---|---|---|---|---|---|---|
| Metric | Crystal system | | | | | Space group | | | | |
| | FA | GBC | RF | DT | SVM | FA | GBC | RF | DT | SVM |
| Accuracy | 0.718 | 0.788 | 0.619 | 0.739 | 0.688 | 0.599 | 0.678 | 0.42 | 0.655 | 0.566 |
| Precision | 0.716 | 0.785 | 0.62 | 0.738 | 0.685 | 0.58 | 0.672 | 0.344 | 0.651 | 0.529 |
| Recall | 0.718 | 0.788 | 0.619 | 0.739 | 0.688 | 0.599 | 0.678 | 0.419 | 0.655 | 0.566 |
| F1-score | 0.716 | 0.786 | 0.578 | 0.738 | 0.671 | 0.587 | 0.674 | 0.327 | 0.652 | 0.532 |
| *FA* fingerprint analysis based on 5-NNs, *GBC* gradient boosting classifier, *RF* random forest, *DT* decision tree, *SVM* support vector machine. | | | | | | | | | | |

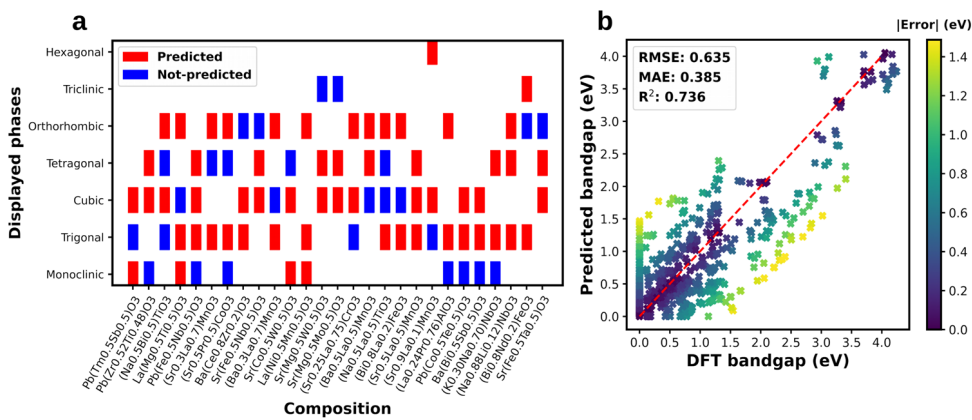

Fig. 5 **Predicting the properties of materials via fingerprint analogies. a** Phase prediction of experimental compositions that can crystallize in three or more phases. We focus only on the crystal system (i.e., the phase transitions occurring within a single crystal system are ignored). The phase is considered to be predicted if at least one of the 5-NNs belong to the crystal system of original composition. **b** Bandgap prediction results for a DFT database of quaternary perovskites[5]. It should be noted that generalized gradient approximation is used for the calculations in ref. [5] that tends to underestimate the bandgap.

### Phase identification and bandgap prediction

Some perovskites such as ferroelectrics display temperature-dependent phase transitions. There can also be coexisting phases at the same temperature[39]. We emphasize that chemical composition is not duplicated in our database, leading to an elimination of additional phases, if there are any. Taking a step back, we examined whether the NNs can reflect phase information of a material that has multiple crystallization phases. We found that if the NNs imply strong relation to two or more different symmetries by means of low euclidean distance (i.e., $S_E > \sim 95\%$), it may indicate a potential phase transition in the material of interest. For example, consider ferroelectric $(Sr_{0.6}Pb_{0.4})TiO_3$ which is not recorded in the ICSD (neither ICSD nor our database contains any record of $Sr - Pb - Ti - O$ system). Four experimental NNs of this composition vote for the cubic system and the remaining NN shows an $S_E$ of 95.7% to the original ferroelectric and votes for the tetragonal system. There is indeed a cubic to tetragonal phase transition at 66 °C in this material[40]. 219 compositions in our experimental database can crystallize in multiple phases (upto 4 crystal systems). In 97 cases, all the crystal systems shown by each material are present in the set of crystal systems of corresponding 5-NNs. We stress that the same experimental fingerprints obtained above are used throughout with no further training. Notably, 33 compositions show 3 or more phases. The stated approach can predict 2 or more phases in 27 out of 33 compositions as depicted in Fig. 5a.

We search for the compositions with no known phase transformations by probing those which are reported to have only one phase (one crystal system) for all alloying ratios within the system. We found that 1283 compositions have no potential phase transformations. Out of these, 573 were predicted to have only one phase as expected. Two phases were predicted for 475 compositions. We conjecture that a considerable number of materials have not been fully characterized and even some of the known phases are not yet recorded in the ICSD. For example, ferroelectric $(Ag_{0.9}Li_{0.1})NbO_3$ favors a trigonal phase at room temperature and displays temperature-induced sequence trigonal → orthorhombic → tetragonal → cubic[41]. However, the ICSD reports only the trigonal phase of this material (as of March 2021). Therefore, it was wrongly labelled to have no phase transformations. We notice that 4-NNs of $(Ag_{0.9}Li_{0.1})NbO_3$ vote for the trigonal system and one votes for the cubic system. This is sensible given the above phases of this material. While capturing the Curie point is too much to ask from this unsupervised approach, it does convey some useful insights on possible phase transitions. We further tested whether the bandgaps of the compositions located in close proximity are similar by predicting the DFT computed bandgaps of 1016 quaternary perovskites reported in ref. [5]. These compositions are fingerprinted and the bandgap is predicted by averaging this figure of 5-DFT-NNs of the considered composition (see Fig. 5b). The mean absolute error of the predictions is 0.385 eV, consistent with existing ML based bandgap prediction studies[12,42]. We further employed several supervised approaches for predicting the bandgap. Although the error rate above is very similar to LOOCV MAE of support vector regressor (0.369 eV), the lowest MAE is achieved by RF regressor (0.17 ± 0.004 eV). Based on

Table 3. Crystal system prediction of perovskites not present in the ICSD using material fingerprints.

| Composition | Experimental symmetry | Predicted symmetry (by fingerprint analogy) | $N_{votes}$ | $S_{\bar{E}}$ (%) |
|---|---|---|---|---|
| $(Sr_{0.7}Pb_{0.3})TiO_3$ | Cubic $Pm\bar{3}m$[40,60] | Cubic $Pm\bar{3}m$ | 4/5 | 95.1 |
| $Ca(Zr_{0.05}Ti_{0.95})O_3$ | Orthorhombic $Pnma$[61] | Orthorhombic $Pbnm$ | 5/5 | 91.5 |
| $(La_{0.9}Pb_{0.1})CoO_3$ | Trigonal $R-3c$[62] | Trigonal $R-3c$ | 4/5 | 78.5 |
| $(La_{0.8}Pb_{0.2})CoO_3$ | Trigonal $R-3c$[62] | Trigonal $R-3c$ | 4/5 | 81.5 |
| $Sr(V_{0.5}Bi_{0.5})O_3$ | Monoclinic[63] | Monoclinic $I2/m$ | 3/5 | 96.2 |
| $(K_{0.5}Nd_{0.5})TiO_3$ | Cubic $Pm\bar{3}m$[64] | Cubic $Pm\bar{3}m$ | 3/5 | 92.7 |
| $(K_{0.5}Sm_{0.5})TiO_3$ | Cubic $Pm\bar{3}m$[64] | Cubic $Pm\bar{3}m$ | 3/5 | 91.2 |
| $Sr(Y_{0.5}Nb_{0.5})O_3$ | Monoclinic $P2_1/n$[65] | Monoclinic $P2_1/n$ | 4/5 | 98.1 |
| $Sr(Hf_{0.5}Ti_{0.5})O_3$ | Cubic $Pm\bar{3}m$[66] | Cubic $Pm\bar{3}m$ | 5/5 | 87.2 |
| $(Sr_{0.8}Ca_{0.2})ZrO_3$ | Orthorhombic $Pnma$[67] | Orthorhombic $Pbcm$ | 4/5 | 94.1 |
| $(Sr_{0.6}Ca_{0.4})ZrO_3$ | Orthorhombic $Pnma$[67] | Orthorhombic $Pbnm$ | 5/5 | 95.8 |
| $(Ba_{0.3}Sr_{0.7})ZrO_3$ | Cubic[68] | Cubic $Pm\bar{3}m$ | 5/5 | 95.5 |
| $(Ba_{0.9}Sr_{0.1})ZrO_3:0.025Eu$ | Cubic[69] | [a]Cubic $Pm\bar{3}m$ | 3/5 | 93.3 |
| $(Sr_{0.6}Pb_{0.4})TiO_3$ | Tetragonal $P4mm$[40] | Cubic $Pm\bar{3}m$ | 4/5 | 95.6 |
| $(Sr_{0.5}Pb_{0.5})TiO_3$ | Tetragonal $P4mm$[40] | Cubic $Pm\bar{3}m$ | 4/5 | 95.7 |
| $(La_{0.95}Sb_{0.05})FeO_3$ | Orthorhombic $Pbnm$[70] | Trigonal $R-3c$ | 4/5 | 91.4 |
| [b]$Ba(Hf_{0.05}Ti_{0.95})O_3$ | Tetragonal $P4mm$ - orthorhombic $Amm2$[39] | Cubic $Pm\bar{3}m$ | 4/5 | 88.7 |

None of the quaternary systems relating to these compositions exist in our database, making them completely unseen to the model. $N_{votes}$ is the vote fraction secured by the predicted crystal system. $S_{\bar{E}}$ is the similarity measure obtained based on the mean of the euclidean distances to 5-NNs.
[a]Prediction refers to $(Ba_{0.9}Sr_{0.1})ZrO_3$ composition without a dopant.
[b]Coexistence of $P4mm$ - $Amm2$ phases is reported.

these predictions, it is safe to assume that geographically close materials likely possess similar properties.

### Analogical discovery of perovskites
The fingerprints of the 10,790 predicted perovskites were extracted at the $\mu$ layer by sending the associated features through the trained VAE network. This brings the candidate compositions and the experimental observations to a common ground, that is the fingerprint space, where unexplored perovskites landing close to a target experimental material are identified as analogous compositions to that material. We first tested the effectiveness of fingerprints for a set of predicted perovskites. During the materials screening stage, our web scraping tool recovered 24 GBC-predicted perovskites from literature that were not available in the ICSD. 17 out of 24 were reported in credible experimental studies that contained crystal structure information. As predicted, all 17 compositions have been experimentally identified as perovskites. By placing them in the fingerprint space and taking the majority vote of 5-experimental-NNs into account, we correctly identified the crystal system of 13 perovskites out of 17 (Table 3). Not only does this underpin the accuracy of our ML prediction, but it also confirms the generalizability of the material fingerprints to unseen data.

The inauguration of analogical materials discovery is driven by the exigent demand for alternatives to some of the existing industrial materials that may be toxic, expensive, unstable or lossy. For example, discovering lead-free ceramics and solar cell materials is a heavily invested specialism[43]. Identifying lead-free promising perovskites that potentially resemble the crystal structure and physiochemical properties of a Pb-based material is achievable by fingerprinting the Pb-composition and capturing the nearest candidates in the fingerprint space. For instance, our analysis suggests that unstudied $Bi(Mg_{0.65}V_{0.35})O_3$ could be a non-toxic substitution for room temperature multiferroic $Pb(Fe_{0.5}Nb_{0.5})O_3$ (PFN). PFN ceramic undergoes several phase transitions, namely, cubic → tetragonal → rhombohedral → monoclinic with decreasing temperature[44]. Rather unexpectedly, 5-NNs to $Bi(Mg_{0.65}V_{0.35})O_3$ evidence the relation to all these phases with tetragonal $P4mm$ receiving two votes and the rest obtaining one vote each. Moreover, $Bi(Sc_{0.2}Co_{0.8})O_3$ and $Bi(Ti_{0.5}Zn_{0.5})O_3$ locate near the relaxor ferroelectric $Pb(Mg_{1/3}Nb_{2/3})O_3$. Note that double perovskite equivalent of $Bi(Ti_{0.5}Zn_{0.5})O_3$ has experimentally been identified as an analogue of $PbTiO_3$[45]. Furthermore, $(K_{0.5}Bi_{0.5})ZrO_3$ shows similarity to the studied ceramic $(Ba_{0.67}Pb_{0.33})TiO_3$. It is well regarded that bismuth-based ceramics are eco-friendly alternatives to lead-rich piezoelectrics, with the downside of usually requiring high-pressure and high-temperature synthesis conditions. Intriguingly, many compositions with bismuth occupying A-site, when fingerprinted, lie close to those containing lead. This implies that our model can rediscover known phenomena even with no domain-specific knowledge explicitly supplied. Full list of suggested alternatives to current selected perovskites is tabulated in Supplementary Table 3.

### DFT validation
One of the major hindrances to modelling unstudied materials with ab initio methods such as DFT is the lack of a priori knowledge about the starting geometry. This is especially challenging for perovskites having previously unknown cation combinations. In the era of experimental records reaching big data status, chances are structurally similar materials to an unstudied material do exist. Not only does material fingerprinting facilitate analogical materials discovery, but it also is readily applicable to find a reasonable experimental structure required to initialize DFT geometry optimization. As a demonstration, we perform DFT calculations to realize the optimized structures of six selected promising perovskites. Tractable computational methods of modelling random solid solutions in conjunction with the DFT framework include the special quasirandom structures (SQS) approach[46], the virtual crystal approximation[47] and the supercell approach that signifies disorder aspects at local level. While the SQS method has been particularly successful in predicting the properties of random solid solutions[48], we note that the supercell approach is significantly faster than SQS generation and the associated groundwork. Since property prediction is not intended here, we opt for the supercell method and use the recent `supercell` program especially designed to construct reflective supercells of solids having mixed/partial site occupancies[49]. It potentially includes SQS systems within its exhaustive search space and implements a combinatorial structure ranking method based on their respective electrostatic (Coulomb) energies. Owing to the importance of conformer search, here, we select the supercell configuration with the lowest Coulomb energy to start geometry optimization. Note that, supercell generation itself could be very computationally demanding depending on the number of combinatorial structures, not to mention the subsequent DFT calculation. Figure 6 elaborates the DFT structure optimization steps of the selected promising perovskite (Ce$_{0.5}$Ag$_{0.5}$)ZrO$_3$. By analyzing the fingerprints, we found that (La$_{0.8}$Al$_{0.2}$)FeO$_3$ orthorhombic *Pbnm*, is the most similar experimental material in our database, to the selected candidate. Hence, it is used as the starting geometry. Table 4 details the structure generation and the optimized lattice constants of the six selected promising perovskites. The structures are fully relaxed, allowing to change the ionic positions, cell volume and cell shape. All the optimized structures preserve the perovskite nature as well as the starting structural integrity with minimal change in lattice parameters, seemingly indicating that the initial geometry found by material fingerprint analogy is a plausible basis.

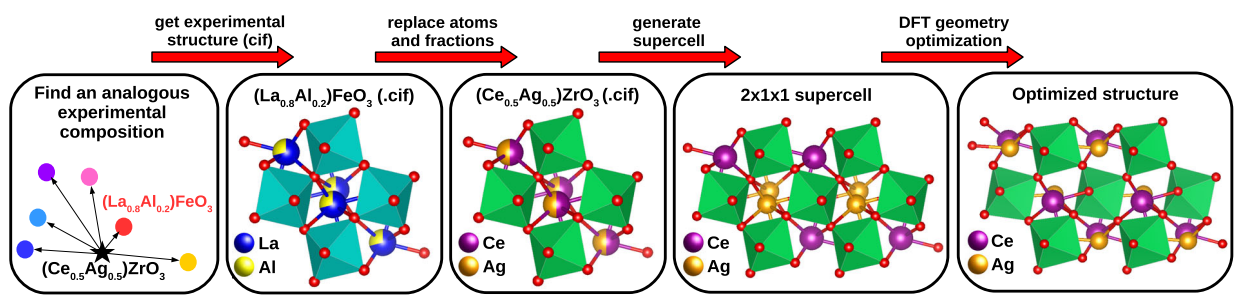

**Fig. 6 DFT modelling workflow of the composition (Ce$_{0.5}$Ag$_{0.5}$)ZrO$_3$.** The initial geometry cif (crystallographic information file) is borrowed from an analogous experimental material (La$_{0.8}$Al$_{0.2}$)FeO$_3$ having an orthorhombic *Pbnm* structure. The atomic species, occupancies and oxidation states were replaced accordingly. A supercell of size 2 × 1 × 1 having the lowest Coulomb energy is selected. DFT optimization of ionic positions, cell volume and cell shape results in an orthorhombic *Pmn*2$_1$ structure, preserving perovskite nature and staring structural attributes.

## DISCUSSION

There are several extensions to the present study worthwhile considering. Virtually all materials in our experimental database are published single phase structures. Collecting non-perovskite materials from the literature or databases essentially means that we miss failed syntheses of perovskites or those having major secondary phases, which are, strictly speaking, non-perovskites too. While this data is not usually published, the possible inclusion of such materials as non-perovskites would enhance the reliability of ML predictions. Another possibility would be to use a transfer learning technique by training a neural network classification model subsequently on a large computational database of perovskites/non-perovskites and available experimental data to mitigate data limitation problem[15,50]. Furthermore, because over 90% of our data has been acquired near room temperature, we did not scrutinize temperature dependence for crystal structure prediction. All our predictions are based only on the composition, assuming ambient conditions. Present study can be extended to include temperature by curating a sufficient database having enough diversity in the temperature field. Due to data limitations, this method is restricted to 66 space groups available in our database. Therefore, the phases outside of these space groups cannot be predicted. This is one of the reasons why we focus on the crystal system, as the data is abundant for all the categories. Moreover, the typical black-box nature of artificial neural networks makes the material fingerprint a black-box too. Investigating the ties of the fingerprint to a physical meaning would make the materials design process more transparent. A relatively large and ideally experimental database of heterogeneous compositions is essential for the proposed framework to succeed. This may be viewed as a constraint to apply our concept in very specific domains where materials data is scarce. We believe the database used in this study is reasonably sized to follow a neural network strategy, and we recommend at least a similar sized database to apply our method in a different context.

In summary, by establishing a comprehensive experimental database of disordered quaternary compositions and involving powerful ensemble ML methods, we have rapidly screened potential perovskites from total feasible compounds. A further series of filtering steps including automated web scraping were imposed to ensure the originality and potential synthesisability of the predicted perovskites. Concurrently, we introduced an unsupervised deep learning strategy to retrieve fingerprints of materials. We elucidated the effectiveness and the educated essence of fingerprints by unravelling hidden salient information such as perovskite formability and underlying crystal structure, that can be used to identify the crystal symmetry of materials. The fingerprints essentially regulate all compositions into a common numerical platform where similar compositions tend to locally stick together. This phenomenon can be capitalized to predict the possible phases and the bandgap of materials. This further led to the conception of analogical materials discovery, that could accelerate the pursuit of finding inexpensive green materials as a replacement for current hazardous materials. We also emphasized the importance of fingerprints to pick up an initial experimental structure required for DFT modelling. Material fingerprint uncovered in this study is not universal, meaning that, it is mutable in a different data context. Nevertheless, due to the versatility of unsupervised techniques, the materials analogy quantification conceptualized here is expected to be valid for other material systems as well. Therefore, we believe the proposed framework has far-reaching implications for accelerated target-driven discovery of materials as well as computational modelling of complex compounds.

## METHODS

### Machine learning

All supervised ML algorithms used in this study, including KNN and GBC are adapted from the `scikit-learn` package[51] in a `python` programming language environment. Gradient boosting algorithms create an ensemble of weak learners, typically by adding DTs over time to produce a powerful predictive model. The learning takes place by applying gradient

**Table 4.** DFT geometry optimization results of 6 selected promising perovskites.

| Candidate Perovskite | Experimental Analogy | Similarity Measure (%) | Supercell Size | Optimized Symmetry | Optimized lattice constants | | | | | |
|---|---|---|---|---|---|---|---|---|---|---|
| | | | | | a | b | c | α | β | γ |
| $(Ce_{0.5}Ag_{0.5})ZrO_3$ | $(La_{0.8}Al_{0.2})FeO_3$ | 90.8 | 2×1×1 40 atoms | Orthorhombic $Pmn2_1$ | 11.257 (+0.232) | 5.901 (+0.374) | 8.132 (+0.33) | 90 | 90 | 90 |
| $Bi(Sc_{0.2}Co_{0.8})O_3$ | $Pb(Mg_{0.33}Nb_{0.67})O_3$ | 98.3 | 1×1×5 25 atoms | Tetragonal $P4/mmm$ | 3.926 (+0.124) | 3.926 (+0.124) | 18.825 (−1.425) | 90 | 90 | 90 |
| $Bi(Mg_{0.65}V_{0.35})O_3$ | $Pb(Fe_{0.5}Nb_{0.5})O_3$ | 92.1 | 1×1×3 15 atoms | Tetragonal $P4mm$ | 3.834 (−0.173) | 3.834 (−0.173) | 13.385 (+1.345) | 90 | 90 | 90 |
| $(K_{0.5}Bi_{0.5})ZrO_3$ | $(Ba_{0.95}Pb_{0.05})TiO_3$ | 92.6 | 1×1×2 10 atoms | Tetragonal $P4mm$ | 4.15 (+0.158) | 4.15 (+0.158) | 8.317 (+0.25) | 90 | 90 | 90 |
| $(K_{0.5}La_{0.5})ZrO_3$ | $(Sr_{0.6}La_{0.4})TiO_3$ | 87.3 | 1×1×2 10 atoms | Tetragonal $P4/mmm$ | 4.139 (+0.005) | 4.139 (+0.005) | 8.375 (+0.012) | 90 | 90 | 90 |
| $Ba(U_{0.5}Sn_{0.5})O_3$* | $Ba_2YbReO_6$ | 90.7 | NA 40 atoms | Cubic $Fm\bar{3}m$ | 8.73 (+0.407) | 8.73 (+0.407) | 8.73 (+0.407) | 90 | 90 | 90 |

Optimized lattice parameters relate to the relaxed supercell. The space group symmetry can be lowered during supercell generation because the initial "mixed" atom gets separated into two atom types. Hence, we substituted the disordered site with one atomic species to determine the optimized symmetry. a, b and c are in angstroms (Å). The change in lattice parameter (in angstroms) compared to experimental analogy is indicated in brackets. α, β and γ are in degrees (°).
*Two probable oxidation state combinations at B-site, ($U^{4+}$, $Sn^{4+}$) and ($U^{6+}$, $Sn^{2+}$) were considered, which converged to almost identical lattice parameters and total energy. ($U^{4+}$, $Sn^{4+}$) is the most probable combination based on ICSD statistics.
NA – not applicable, experimental structure is used as it is.

descent algorithm to optimizing a differentiable cost function, usually logarithmic loss for classification tasks. The hyper-parameters are external to the ML model and should be tuned in advance to control the training process and achieve better predictive power. We tuned the hyper-parameters of the models (learning rate and No. of estimators in the case of GBC) using tenfold CV and grid search method. In k-fold CV, the database is randomly partitioned into k equal size subsets. The ML model is trained by combining k-1 of them and validated on the remaining subset. This is repeated k times such that each data-point in the original database gets to be in the test set exactly once. LOOCV is the extreme version of k-fold CV where k is equal to the number of samples in the database. Material features were acquired and analyzed using `pymatgen`[34] and `mendeleev`[52] python packages. The computer codes required to train the models and reproduce the results presented in this paper may be found online at https://github.com/ihalage/analogmat.

### VAE architecture and training
The VAE model is implemented in `python` using `Keras` deep learning framework[53] with `TensorFlow` backend[54]. The depth of the VAE neural network and the dimension of the latent vector should be carefully controlled to obtain meaningful representations. A high depth (i.e., more parameters) means low reconstruction loss, but this often comes with the risk of data memorization rather than learning anything useful. High dimensional latent space can also give rise to this problem. We found that, using a significantly low dimension for the latent vector not only forces the encoder to find near optimal compression due to the added pressure by high reconstruction loss, but it also facilitates convenient data visualization without incorporating other dimensionality reduction algorithms. After some extensive heuristic tuning, we eventually constructed the VAE with five hidden layers with 2D $\mu$, $\sigma$ and latent vectors. The remaining four hidden layers have 128, 64, 32, and 64 neurons, chronologically. The weights were initialized using Glorot uniform initializer. Rectified linear unit is used as the activation function of the hidden layers and linear activation is used for the output layer. $L_2$ regularization is applied in hidden layers to avoid the overfitting phenomenon. Kullback–Leibler divergence (KL loss)[55] and mean squared error (MSE) were used additively as the loss function, which was optimized using Adam optimizer with a learning rate of 0.001. The exponential decay parameters $\beta_1$, $\beta_2$ and the stability constant $\epsilon$ were set to 0.9, 0.999 and $10^{-7}$, respectively. The model is trained on an RTX 2080 Ti graphics processing unit (GPU) with a batch size of 32 and the validation data (20%) loss was monitored for an early-stopping method to prevent possible data memorization.

### Density functional theory
Spin-polarized DFT calculations were performed with projector augmented wave pseudopotentials[56] using the generalized gradient approximation as implemented in the Vienna Ab initio Simulation Package (VASP)[57]. We used perdew–burke–ernzerhof exchange-correlation functional for the present calculation. The Brillouin zone was sampled with a 10×10×10 Monkhorst-Pack k-point grid. Cutoff energy for the plane wave basis set was set to 520 eV. The ionic positions, cell volume and cell shape were allowed to change during the structure relaxation until reaching an energy convergence threshold of $2 \times 10^{-6}$ eV and net atomic force less than 0.01 eV/Å on any atom. The solid solutions were approximated with a supercell having the lowest Coulomb energy as generated by the `supercell` package[49]. The supercell dimensions used in this study are sufficient to demonstrate the geometry optimization process, however, larger supercells or a SQS based method is recommended if further calculation of physical properties is intended. The crystal structures were analyzed using `pymatgen` and visualized with `VESTA`[58]. We used the GPU port of VASP[59] to run the calculations on an RTX 2080 Ti GPU.

### DATA AVAILABILITY
The experimental database is provided in Supplementary Data 1. Predicted compositions having over 98% perovskite likelihood are provided in Supplementary Data 2 along with their respective five most similar experimental materials. All other data needed to evaluate the findings of this study are present in the paper and/or the Supplementary Information.

## CODE AVAILABILITY

The codes required to reproduce the results of this study are publicly available at https://github.com/ihalage/analogmat.

## ACKNOWLEDGEMENTS
The authors thank Sachini Amararathna for collecting the experimental data analyzed in this study. We thank M.J. Reece for stimulating discussions on disordered materials and their attributes. The authors acknowledge funding received by The Institution of Engineering and Technology (IET) under the AF Harvey Research Prize. This work is supported in part by Engineering and Physical Sciences Research Council (EPSRC) ANIMATE grant (No. EP/R035393/1).

## AUTHOR CONTRIBUTIONS
A.I. conceived the idea, performed ML and DFT analysis and wrote the paper. Y.H. directed and coordinated the research. All authors discussed the results and reviewed the paper.

## COMPETING INTERESTS
The authors declare no competing interests.

## ADDITIONAL INFORMATION
**Supplementary information** The online version contains supplementary material available at https://doi.org/10.1038/s41524-021-00536-2.

**Correspondence** and requests for materials should be addressed to Y.H.

**Reprints and permission information** is available at http://www.nature.com/reprints

**Publisher's note** Springer Nature remains neutral with regard to jurisdictional claims in published maps and institutional affiliations.